\documentclass[prl,twocolumn,showpacs,preprintnumbers,amsmath,amssymb]{revtex4}
\usepackage{graphicx}

\begin{document}
\title{Universal features of electron-phonon interactions in atomic wires}
\author{L. de la Vega$^1$, A. Mart\'{\i}n-Rodero$^1$, N. Agra\"{\i}t$^2$
and A. Levy Yeyati$^1$}
\affiliation{$^1$Departamento de F\'{i}sica Te\'{orica} de la Materia
Condensada CV}
\affiliation{$^2$Departamento de F\'{\i}sica de la Materia Condensada CIII,
 Universidad Aut\'{o}noma de Madrid, E-28049
Madrid, Spain}
\begin{abstract}
The effect of electron-phonon interactions in the
conductance through metallic atomic wires is theoretically analyzed.
The proposed model allows to consider an atomic size region
electrically and mechanically coupled to bulk electrodes.
We show that under rather general conditions the features due to
electron-phonon coupling are described by universal functions of the system
transmission coefficients. It is predicted that the reduction of the conductance
due to electron-phonon coupling which is observed close to perfect transmission
should evolve into an enhancement at low transmission. This crossover can be
understood in a transparent way as arising from the competition between
elastic and inelastic processes.
\end{abstract}
\maketitle

\section{Introduction}

Metallic nanowires, including atomic contacts and atomic chains have become
ideal systems for testing predictions of quantum transport theory \cite{review}.
Many transport properties in the normal state like shot-noise \cite{Cron01},
conductance fluctuations \cite{Ludoph00} and others in the
superconducting state \cite{Scheer97-98} have been measured
with a remarkable agreement with the theoretical predictions \cite{review}.
A key ingredient in this analysis is provided by the set of transmission coefficients
${\tau_n}$ (the so-called
mesoscopic PIN-code), in terms of which a generic transport property $F(V)$,
where $V$ is the bias voltage,
can be written as $F(V) = \sum_n f(V,\tau_n)$, $f(V,\tau)$ being the
contribution of a single conduction channel \cite{Scheer97-98}. 
The validity of this decomposition relies on the usually
large difference in energy scales between the electronic excitations involved,
$\sim eV$, and the typical scale $W_{el}$ 
for the variation of the normal density
of states. This one-electron description applies as far as charging effects 
can be neglected due to the strong coupling of the electronic states of the
atomic region with the metallic electrodes. 

Recently there has been much interest in the formation of metallic atomic
chains using STM and break junction techniques \cite{Yanson99} allowing
the possibility to study electron transport through these nearly ideal
one-dimensional systems.
In particular, experiments in Au chains show unambiguously the presence of
features associated to a quasi one-dimensional phonon spectrum \cite{Agrait}.
The experimental conductance exhibits tiny steps at voltages
corresponding to vibrational modes of the finite suspended chain.
A number of theoretical papers have addressed the explanation of these basic
features \cite{Todorov,Frederiksen,Tossati}.
It could be interesting to investigate whether the
type of description commented above in terms of the system transmission
coefficients could be valid to analyze the effects of
electron-phonon (EPH) interactions, which is reasonable as the  
relevant phonon energies $\hbar \omega \sim 10$ meV are 
still small compared to the typical $W_{el} \sim 0.1-1$ eV in metallic 
atomic conductors \cite{Cuevas98}.
Starting from a microscopic model and performing
reasonable simplifying assumptions we show in this work that the main
features due to EPH coupling in a single-channel
conductor can be described by a universal function of its transmission $\tau$.
The sign of the conductance correction, which is negative for large
transmission, evolves into a positive correction for low transmission.
This behavior can be understood as a competition between
elastic and inelastic processes. We show that these predictions apply
naturally to the case of atomic chains of arbitrary length.
\begin{figure}
\includegraphics[width=0.8\columnwidth]{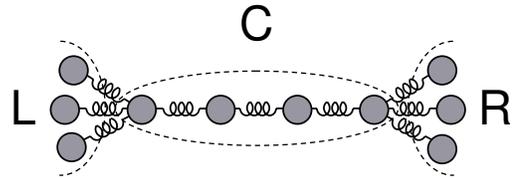}
\caption{Schematic representation of an atomic wire
suspended between metallic electrodes. For the theoretical description
the system is decomposed into L (left), R (right) and C (central) regions.
The labels 1 and $N$ denote the outermost sites of the wire.}
\label{sistema}
\end{figure}

\section{Theoretical Model}

As a starting point we consider a tight-binding model for describing the electronic properties which is well adapted to spatially inhomogeneous
systems like atomic size conductors \cite{Cuevas98,mas}. This approach can
be straightforwardly extended to incorporate
the phonon degrees of freedom \cite{Todorov}. We consider the type of
geometry schematically depicted in Fig. \ref{sistema} consisting of a left and right
electrode electrically and
mechanically coupled to a central atomic size region.
The model allows in principle to incorporate not only the phonons of the isolated central
region (as done in most previous works \cite{Todorov,Frederiksen,Tossati})
but also to take into account the effect of the electrodes in their dynamics.
Within this approach the electronic Hamiltonian can be
written as $\hat{H}_{el} = \sum_{ij,\sigma} h_{ij} \hat{c}_{i\sigma}^{\dagger}\hat{c}_{j\sigma}$, where
$\hat{c}^{\dagger}_{i\sigma}$ creates an electron in site $i$ with spin $\sigma$ ($i,j$ run over sites
of the whole system); and equivalently for the phonon degrees of freedom we have $\hat{H}_{ph} = \sum_i \hat{p}_i^2/2M + \sum_{ij} \hat{u}_{i}
A_{ij} \hat{u}_{j}$, where the atomic displacements $\hat{u}_i$ and the
corresponding momenta $\hat{p}_i$
satisfy the quantization relation $[\hat{u}_j,\hat{p}_k]=i\hbar \delta_{jk}$.
Both the electronic Hamiltonian and the dynamical
matrix ${\bf A}$ can be decomposed as $\hat{H}= \hat{H}_L + \hat{H}_R + \hat{H}_C + \hat{V}_{LC} + \hat{V}_{CR}$,
describing respectively the isolated $L$, $R$ and $C$ regions and the coupling
between the central region and
the electrodes. Finally, the EPH coupling
$\hat{H}_{e-ph} = \sum_{ij} \lambda_{ij} (\hat{u}_i-\hat{u}_j)
\hat{c}^{\dagger}_j\hat{c}_i$ is obtained
by expanding the electronic Hamiltonian to first order in the lattice
displacements \cite{Todorov}.
We shall consider the simplest version of this tight-binding model where only the hopping terms between neighboring sites are non-zero and thus the
EPH coupling constants
$|\lambda_{ij}|=\partial h_{ij}/\partial r$ are non-local and short ranged.

The transport properties of this model are conveniently
calculated in terms of
non-equilibrium Green function techniques. Using the electronic
propagators in Keldysh space $G_{ij}^{\alpha\beta}(t,t')$, where
$\alpha,\beta\equiv+,-$
denote the two branches of the Keldysh contour \cite{Keldysh},
the current between two sites $i,j$ can be written as
\begin{equation}
I_{ij} = \frac{2e}{\hbar} h_{ij} \left(G^{+-}_{ij}(t,t) - G^{+-}_{ji}(t,t)\right).
\end{equation}
The EPH interaction is included to second order in perturbation
theory, which is appropriate in the metallic case \cite{Migdal}.
The corresponding
self-energy insertions can be written in terms of the unperturbed
propagators as
\begin{equation}
\Sigma^{\alpha\beta}_{ij}(\omega) = i \alpha\beta
\int \frac{d\omega'}{2\pi}
\sum_{kl} \lambda_{ik}\lambda_{lj}
\bar{D}^{0,\alpha\beta}_{ik,lj}(\omega-\omega')
G^{0,\alpha\beta}_{kl}(\omega'),
\end{equation}
where
$\bar{D}^{0,\alpha\beta}_{ik,lj}
= D^{0,\alpha\beta}_{il} - D^{0,\alpha\beta}_{ij}
- D^{0,\alpha\beta}_{kl} + D^{0,\alpha\beta}_{kj}$,
with $D^{0,\alpha\beta}_{ij}$ denoting the
phonon propagators. If the phonons are assumed to be in thermal 
equilibrium \cite{comment-eq},
the frequency dependent propagators in Keldysh space are related to the
advanced and retarded ones by
$D^{0,\alpha\beta}_{ij} = (n_B+\delta_{\beta,+}) D^{0,r}_{ij} -
(n_B+\delta_{\alpha,-}) D^{0,a}_{ij}$,
where $n_B$ is the Bose-Einstein distribution. On the other hand, the
retarded and advanced propagators
are formally given by ${\bf D}^{0,r,a}(\omega) =
\left[(\omega\pm i\eta)^2 - {\bf A} \right]^{-1}$ and can be evaluated using
similar techniques as for their electronic counterparts \cite{economou}.
Up to this level of approximation, the correction to the current
due to EPH coupling in the chain geometry of Fig. \ref{sistema}
can be decomposed into elastic and inelastic contributions, ie. $\delta I = \delta I_{el} + \delta I_{in}$ \cite{Caroli}, where
\begin{widetext}
\begin{eqnarray}
\delta I_{el} & = & \frac{16e}{h}
\Gamma_L \Gamma_R \int d\omega \sum_{ij} \mbox{Re}\left(G^{0,r}_{1i}
\Sigma^{r}_{ij} G^{0,r}_{jN} G^{0,a}_{N1} \right)
(f_L(\omega) - f_R(\omega)) \label{full-el} \\
\delta I_{in} & = & -\frac{8ei}{h}
\Gamma_L \Gamma_R \int d\omega \int \frac{d\omega'}{2\pi}
\sum_{ijkl} \lambda_{ik} \lambda_{lj} G^{0,r}_{1i}(\omega)G^{0,r}_{kN}(\omega')
G^{0,a}_{Nl}(\omega')G^{0,a}_{j1}(\omega) \nonumber \\
& & \times \left[\bar{D}^{0,-+}_{ik,lj}(\omega-\omega')
f_L(\omega)\left(1-f_R(\omega')\right) -
\bar{D}^{0,+-}_{ik,lj}(\omega-\omega')
f_R(\omega')\left(1-f_L(\omega)\right) \right],
\label{full-in}
\end{eqnarray}
\end{widetext}
$\Gamma_{L,R}$ and $f_{L,R}$ being the tunneling rates coupling the chain
to the leads (assumed to be energy independent) and the corresponding Fermi
distributions.
Within this decomposition the inelastic contribution corresponds to the
transfer of an electron
with the real emission (or absorption at finite temperature) of a phonon mode;
while in the elastic process the interaction with phonons induces
a transition into an intermediate virtual state resulting in a
renormalization of the transmission.
\begin{figure}
\includegraphics[width=0.9\columnwidth]{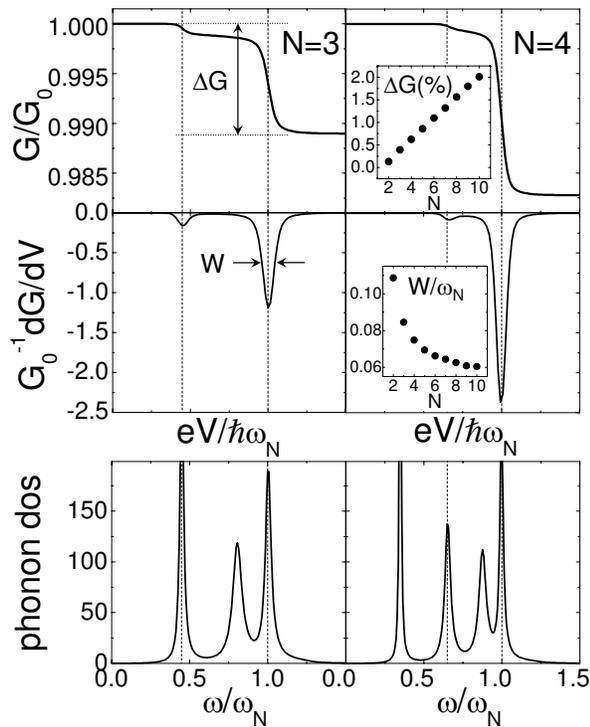}
\caption{Features due to EPH coupling in the conductance
for the ideal chain geometry containing 3 and 4 atoms. The insets in the right upper and middle panels show the scaling of the total conductance step
$\Delta G$ and its width $W$ with the number of atoms $N$.
The correlation of these features with the phonon spectrum is
illustrated by the included average phonon density of states within
the chain in the lower panels. $G_0$ denotes the quantum of conductance
$2e^2/h$.}
\label{idealchain}
\end{figure}

\section{Results}

\begin{figure}
\includegraphics[width=\columnwidth]{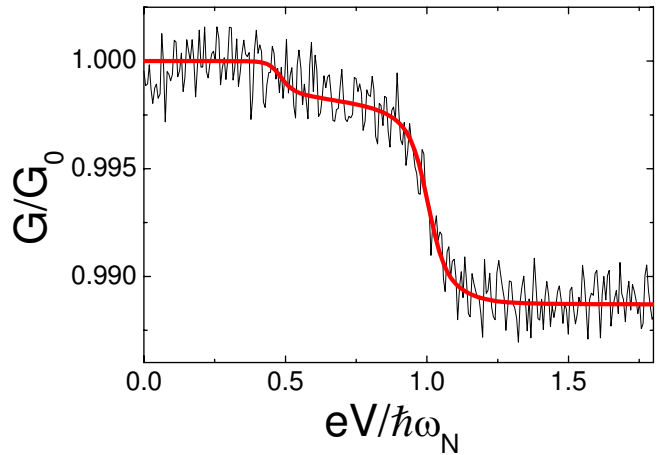}
\caption{(color online) Comparison between 
the full model calculations and the experimental measurements
for the conductance of Au atomic chains. The experimental data
correspond to a typical curve from the experiments described in 
Ref. \cite{Agrait}, while the theoretical curve has been obtained
for the case $N=3$ using the set of parameters discussed in the text.}
\label{compara-exp}
\end{figure}

Let us first analyze the effect of EPH interactions in the
conductance for the ideal chain geometry.
The main features of this case at zero temperature
are illustrated in Fig. \ref{idealchain}.
The ratio between the first neighbors EPH
coupling constant $\lambda$ and the hopping term $t$ within the chain 
is taken to be $7 \times 10^{-3}$ while the maximum phonon mode of
the uncoupled chain is taken as $\hbar \omega_N \sim 0.03t$ 
(i.e. $\hbar \omega_N/t \ll 1$)
in order to reproduce the typical features observed
experimentally for Au chains \cite{Agrait}. The comparison with the
experimental results is illustrated in Fig. \ref{compara-exp} which
will be discussed below.
As can be observed in Fig. \ref{idealchain}, the conductance exhibits
small steps which can be associated with the onset of inelastic emission of
longitudinal phonons. The more pronounced structure is associated with
the highest phonon mode of the chain but smaller features related to
lower modes also appear. In this zero temperature limit it is possible to
correlate the width of the steps with the corresponding width of the
resonances observed in the phonon density of states of the chain (see
lower panels of Fig. \ref{idealchain}). This width is
in turn controlled by the mechanical coupling of the chain with the bulk electrodes, thus providing and intrinsic width of the steps in the
conductance \cite{comment}.
Notice that only steps associated with phonon modes having a definite
parity (even or odd) are observed for a given number of atoms in the chain.
The total step size is found to follow simple scaling laws with
the EPH coupling $\lambda$ and the highest phonon frequency
of the chain $\omega_N$, i.e $\delta G \sim \lambda^2/\omega_N$,
while it increases roughly linearly with $N$ as illustrated in the upper 
right inset of Fig. \ref{idealchain}, in agreement with the experimental 
observations \cite{Agrait}. Physically, this behavior arises from the 
increase of the interaction probability with increasing chain length and
from the increase of the vibration amplitude for decreasing frequency.
The width of the step is also found to scale simply as
$\sim 1/N$ (see inset in right middle panel of Fig. \ref{idealchain}).
As can be observed in Fig.\ref{compara-exp}.
there is a  good agreement between our model calculations
and the typical experimental results of Ref. \cite{Agrait}. 
The theoretical curve corresponds to a short chain ($N=3$), and exhibits 
a small step at low bias in addition to the main conductance step. 
In spite of the experimental noise this
smaller feature can also be distinguished in the experimental curve.

\begin{figure}
\includegraphics[width=\columnwidth]{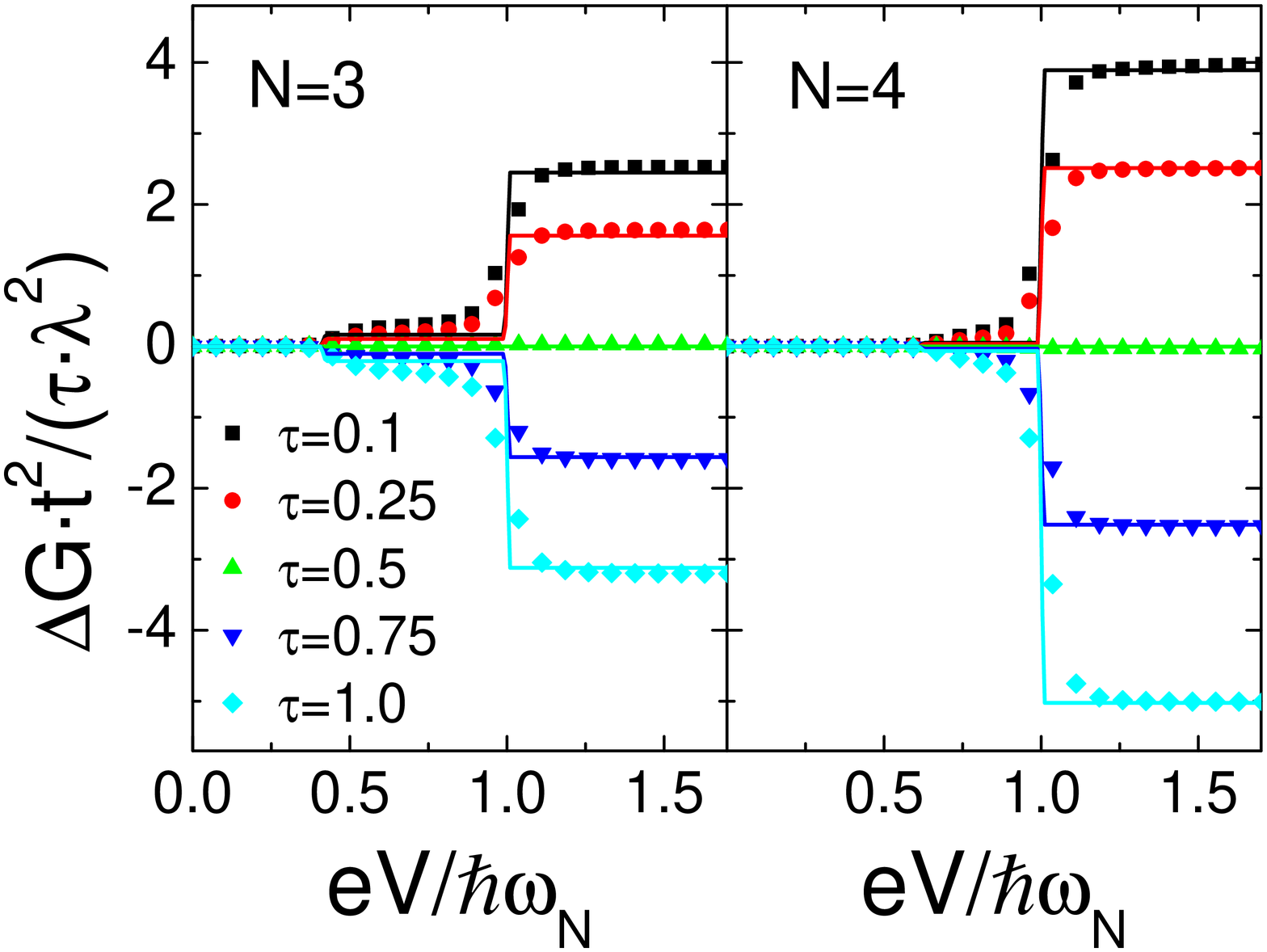}
\caption{(color online) Evolution of the conductance steps as a function of the
transmission for an atomic chain with $N=3$ and 4. The symbols
correspond to the full calculation while the full lines were obtained
using Eqs. (\ref{wideband-el},\ref{wideband-in}).}
\label{ephvstau}
\end{figure}

We have also analyzed deviations from the ideal chain geometry, in
particular we have studied the effect of a zig-zag configuration, taking into account the appearance of
transversal modes in the chain. We have found that these effects do not modify
substantially the overall behavior discussed above. The robustness of these
results and
the simplicity of the scaling behavior found for the linear chain suggest
that the theory could be reduced to a minimal model keeping only the essential
ingredients. 
As we show below this is still the case when the bare channel
transmission is reduced with respect to one.
Thus, assuming that the typical phonon energies are small on the scale 
$W_{el}$,
the energy dependence of the unperturbed electron propagators can be neglected.
One can also relate these propagators to the scattering properties of the
system at the Fermi level by means of the relations \cite{Fisher-Lee}
\begin{equation}
\hat{s} = \left(\begin{array}{cc} r_{L} & t_{LR} \\ t_{RL} & r_{R} \end{array}
\right) = {\bf I} - 2i \left(\begin{array}{cc} \Gamma_L G^{0,r}_{11} & \sqrt{\Gamma_L\Gamma_R} G^{0,r}_{1N} \\
\sqrt{\Gamma_L\Gamma_R} G^{0,r}_{N1} & \Gamma_R G^{0,r}_{NN} \end{array} \right).
\end{equation}
These coefficients are related to the normal transmission $\tau$ by
$\tau = |t_{LR}|^2 = |t_{RL}|^2 = 1 - |r_{L,R}|^2$.
Notice that the transmission can be reduced either due to asymmetries in the
coupling to the electrodes or to the presence of an impurity atom
within the chain.
On the other hand, if one is interested in the total size of the steps
and not in their width, the expressions can be greatly simplified 
by taking the limit of negligible broadening of the chain phonon 
states while maintaining their thermal population. In this limit it is 
possible to factorize the corrections
to the current into a purely electronic factor and another one associated
with the coupling to the phonon system. Up to a uniform background correction
$\delta G_b$, which does not exhibit structure
at the phonon frequencies, one obtains from Eqs.
(\ref{full-el},\ref{full-in}) simple analytical expressions for the conductance
which for a half-filled band and at zero temperature reduce to
\begin{eqnarray}
\delta G_{el} - \delta G_b = -\frac{2e^2}{h} \tau^2 \frac{\lambda^2}{t^2}
\sum_{r=1}^N F_{eph,r} \theta\left(|eV| - \hbar\omega_r\right)
\label{wideband-el} \\
\delta G_{in} = \frac{2e^2}{h} \tau \left(1-\tau\right) \frac{\lambda^2}{t^2}
\sum_{r=1}^N F_{eph,r} \theta\left(|eV| - \hbar\omega_r\right),
\label{wideband-in}
\end{eqnarray}
where $F_{eph,r} = |\sum_{j=1}^{N-1} (-1)^{j}(a_{j,r}-a_{j+1,r})|^2$,
with $a_{j,r}$ denoting the amplitude of the chain atom at site
$j$ in the mode of frequency $\omega_r$
(normalization $\sum_{j=1}^N a_{j,r}^2 = \hbar/M\omega_r$).
It is interesting to observe that these two contributions are universal
functions of the transmission through the chain. Notice that the inelastic
contribution is always positive while the elastic one yields always a
reduction of the total current.
On the other hand, while the elastic contribution increases monotonously with $\tau$, the inelastic correction behaves as $\tau(1-\tau)$ and thus
vanishes in the limit of perfect transmission as in the case of 
shot noise in a quantum point contact \cite{Khlus}.
As a result of the competition between the elastic and the inelastic parts,
the total correction exhibits a change of sign as a function of the
transmission. In this simplified model
the step in the current at the phonon frequencies scales as $\tau(1-2\tau)$,
i.e. the change in the sign of the correction occurs for $\tau=1/2$.
\begin{figure}
\includegraphics[width=0.6\columnwidth]{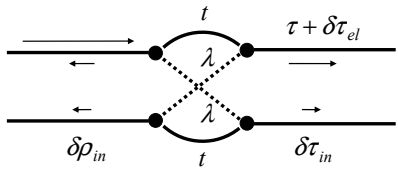}
\caption{Schematic representation of a minimal model accounting for
the transition between negative and positive conductance step as a function
of the transmission based in the mapping proposed in Refs. \cite{Bonca}. A single phonon mode between two sites is considered. The two channels
correspond to the absence or to the presence of an excited phonon.
The arrows indicate all possible paths for an incident electron:
elastic and inelastic transmission and reflection.}
\label{esquema}
\end{figure}
This type of behavior can also be qualitatively obtained by means of a
mapping of the EPH problem into a coherent multichannel
scattering model, as proposed by
several authors \cite{Bonca}. For the simplest case of a single localized mode
between two atoms in which an incoming electron can be transmitted either
elastically or inelastically, as schematically depicted in Fig. \ref{esquema},
this model would predict $\delta \tau_{el} \approx -(\lambda/t)^2
\tau (1-r)(1+2r)$, where $r=|r_{L,R}|=\sqrt{1-\tau}$, and
$\delta \tau_{in} \approx (\lambda/t)^2 \tau(1-\tau)$,
i.e. a change in the sign of the correction for $\tau \sim 0.41$;
while the inelastic reflection is given by
$\delta\rho_{in} = (\lambda/t)^2 \tau^2$.
Thus in the $\tau \rightarrow 1$ ($\tau \rightarrow 0$) limit the
inelastic transitions correspond mainly to back (forward) scattering.
Although this extremely simple model accounts for the overall behavior,
the absence of a correct inclusion of the
restrictions imposed by the Pauli principle is responsible for the
discrepancies with
the correct result (Eqs. (\ref{wideband-el},\ref{wideband-in})).
It is worth noticing that the factor $F_{eph,r}$ in Eqs.
(\ref{wideband-el},\ref{wideband-in}) shows explicitly the selection rules
of modes with a definite parity. For the highest
frequency mode one has
$a_{i,N} \approx -a_{i+1,N}$ which produces a maximum size for the
corresponding step. It is also straightforward to show that the size of
this step increases roughly linearly with $N$.
The evolution of the conductance steps as a function of the transmission 
predicted by Eqs.(\ref{wideband-el},\ref{wideband-in}) is compared in
Fig. \ref{ephvstau} with the results of the full calculation.
As can be noticed the behavior of the full calculation is reproduced
except for the smearing of the steps due to the intrinsic width of the
phonon modes.

We have also analyzed in detail the validity of the simplified model
leading to Eqs.(\ref{wideband-el},\ref{wideband-in})
for significant deviations with respect to the half-filled band case.
One obtains corrections to these equations which are of order 
$(\epsilon/W_{el})^2$,
where $\epsilon$ measures the shift of the band with respect to
half-filling. We have checked numerically that these corrections
become of importance only in the limit $(\epsilon/W_{el}) \sim 1$, 
i.e. close to the nearly empty or full cases.

\section{Conclusions}

We have presented a theoretical analysis of EPH interactions
in atomic size conductors. Our description allows to incorporate in an
equal footing the effects due to electrical and mechanical coupling with
the electrodes. We have shown that in the typical conditions corresponding
to metallic atomic chains the main features of EPH interactions
can be accounted for by simple analytical
expressions factorized into a purely electronic part and another
describing the EPH coupling. While the electronic part
can be written in terms of the bare transmission, the phonon part
accounts for the selection rules and the dependence on the length observed
for the case of an ideal chain.
A prediction which deserves further experimental investigation
concerns the inversion of the conductance steps for a single channel
at $\tau \sim 1/2$.
Atomic Au chains containing S or O impurities which can reduce
substantially its transmission as well as atomic contacts exhibiting
partially open channels like for Al \cite{Cuevas98}
are promising candidates to test these predictions.

{\it Note added:} while completing this manuscript we became aware
of related work by M. Paulsson et al. (cond-mat/0505473)
and J.K. Viljas et al.(cond-mat/0508470). The
authors would like to thank financial support from Spanish MEC
under contracts BMF2001-0150 and MAT2004-03069, and MAT2002-11982E
through the SONS program of the ESF which is also supported by the
EC, 6th Framework Program. Fruitful discussions with J.C. Cuevas
and J.M. van Ruitenbeek are acknowledged.


\begin{thebibliography}{}
\bibitem{comment-eq} This is consistent with the assumption of good 
coupling with the bulk phonons. Deviations with respect to the equilibrium
distribution are not relevant for the present discussion 
and will be analyzed elsewhere.
\bibitem{comment} Other possible mechanisms providing an intrinsic width of
the phonon spectrum like the emission of
electron-hole pairs are expected to give a much smaller contribution.
\end{thebibliography}
\end{document}